\documentclass{iau} 
\usepackage{graphicx}

\title[PSR~B1828$-$11] 
{Magnetospheric Switching in PSR~B1828$-$11 }

\author[I.~H. Stairs et al.]   
{I.~H. Stairs,$^{1}$
A.~G. Lyne,$^{2}$
M. Kramer,$^{3,2}$
B.~W. Stappers, $^{2}$\\
J. van Leeuwen,$^{4,5}$
A. Tung,$^{1,6}$
R. N. Manchester,$^{7}$, G. B. Hobbs,$^{7}$\\
D. R. Lorimer,$^{8,9}$
A. Melatos$^{10}$
} 

\affiliation{
$^{1}$Dept. of Physics and Astronomy, University of British Columbia,
6224 Agricultural Road, Vancouver, B.C., V6T 1Z1, Canada\\ email: {\tt stairs@astro.ubc.ca}\\[\affilskip]
$^{2}$Jodrell Bank Centre for Astrophysics, School of Physics and Astronomy, University of Manchester, Manchester, M13 9PL, UK\\[\affilskip]
$^{3}$Max-Planck-Institut f\"ur Radioastronomie, Auf dem H\"ugel 69, D-53121 Bonn, Germany\\[\affilskip]
$^{4}$ASTRON, The Netherlands Institute for Radio Astronomy, Postbus 2, 7990 AA, Dwingeloo, The Netherlands\\[\affilskip]
$^{5}$Anton Pannekoek Institute for Astronomy, Univ. of Amsterdam, Science Park 904, 1098 XH Amsterdam, The Netherlands\\[\affilskip]
$^{6}$Teacher Education Office, University of British Columbia, 2125 Main Mall, Vancouver, BC V6T 1Z4, Canada\\[\affilskip]
$^{7}$CSIRO Astronomy and Space Science, Marsfield NSW 2122, Australia\\[\affilskip]
$^{8}$Department of Physics and Astronomy, West Virginia University, PO Box 6315, Morgantown, WV 26506, USA\\[\affilskip]
$^{9}$Center for Gravitational Waves and Cosmology, Chestnut Ridge Research Building, Morgantown, WV 26505, USA\\[\affilskip]
$^{10}$School of Physics, University of Melbourne, Parkville VIC 3010, Australia
}

\pubyear{2017}
\volume{337}  
\jname{Pulsar Astrophysics -­ The Next 50 Years}
\editors{P. Weltevrede, B.B.P. Perera, L. Levin Preston \& S. Sanidas, eds.}

\begin{document}

\maketitle

\begin{abstract}
PSR B1828$-$11 is a young pulsar once thought to be undergoing free
precession and recently found instead to be switching magnetospheric
states in tandem with spin-down changes.  Here we show the two extreme
states of the mode-changing found for this pulsar and comment briefly
on its interpretation.
\end{abstract}

\firstsection 
\section{Introduction}

PSR~B1828$-$11 is young pulsar with a spin period of 0.405\,s and DM
of about 160\,pc\,cm$^{-3}$ (\cite{clj+92}).  Years of routine
observations with the Jodrell Bank Observatory showed that its period
derivative varied with an approximately dual-sinusoidal pattern with
periods of about 500 and 250 days.  At the same time, the profile was
seen to vary between ``wide'' and ``narrow'' states, with the average
profile shape following the same pattern as $\dot P$.  These phenomena
were initially interpreted as evidence for free precession
(\cite{sls00}) and an abundant literature sprang up (e.g.,
\cite{lin06}) attempting to understand such precession given the
standard picture of the superfluid neutron-star interior with vortices
pinned to the crust (e.g., \cite{swc99}).

The recognition of a difference of 50\% in spin-down rate in
PSR~B1931+24 when that intermittent pulsar was in one of its
weeks-long off states (\cite{klo+06}) led to a new interpretation for
these and other pulsars: namely that the changes in $\dot P$ were
related to abrupt switches in the profile state, a phenomenon labeled
``magnetospheric switching'' (\cite{lhk+10}).  PSR B1828$-$11 was also
noted to show evidence of rapid profile changes, as in mode-changing
pulsars.  Meanwhile, other authors continued to develop alternative
models for the effects seen in PSR~B1828$-$11, including free
precession combined with magnetospheric switching (\cite{jon12}) and
non-radial oscillations (\cite{rmt11}).  Seymour \& Lorimer (2013)
\nocite{sl13} found evidence that the $\dot{P}$ variations are
consistent with behaviour seen in low-dimensional chaotic
systems. Their analysis suggests that the variations could be produced
by a coupled system of three differential equations with $\dot{P}$ as
one of the “governing variables” controlling the changes in the
pulsar.

\begin{figure}[t]
\begin{center}
 \includegraphics[width=0.9\linewidth]{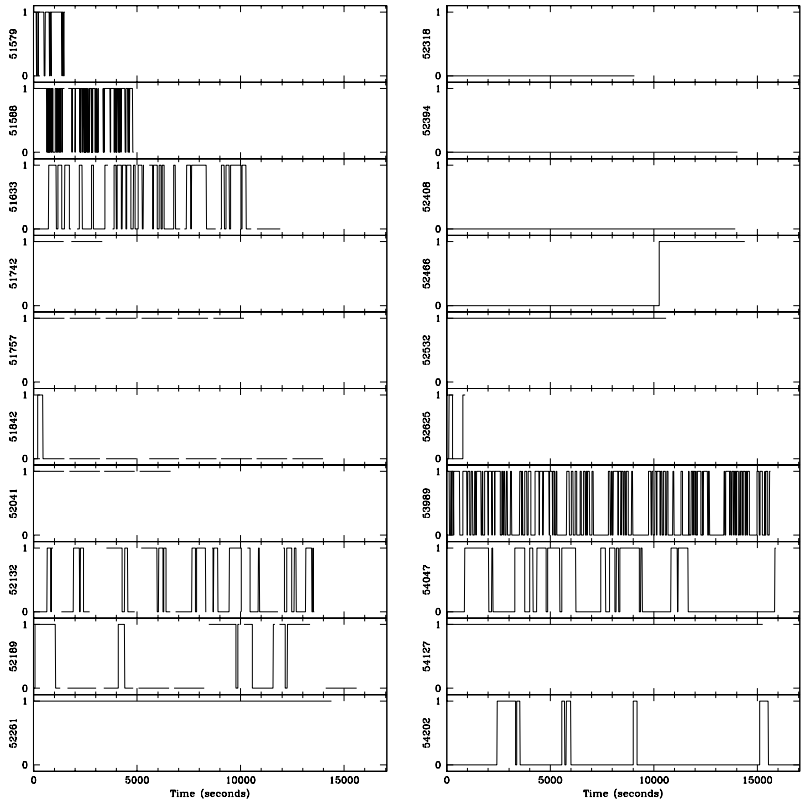} 
 \caption{Narrow (1) and wide (0) states identified for
   the long data sets at different observing epochs.  The last 4 epochs were observed with the GBT;
   the rest with Parkes.}
   \label{states}
\end{center}
\end{figure}

\section{Data and Results}

We obtained long (generally multi-hour) 1400-MHz observations of
PSR~B1828$-$11 on 20 occasions, sampling multiple cycles of the
$\dot P$ variations as determined by Lyne et al.\
(2010). \nocite{lhk+10} On 16 occasions we used the Parkes telescope,
employing the 2 $\times$ 512 $\times$ 0.5 MHz 1-bit filterbank with
samples every 0.25\,ms. On the last 4 occasions we used the Green Bank
Telescope (GBT), employing the BCPM (\cite{bdz+97}) with 96 1-MHz
frequency channels with 72-$\mu$s sampling.  Each observation was
folded with 10-second sub-integrations using {\tt dspsr} (\cite{vb11})
and wide and narrow profiles were identified by eye using viewing
programs from the {\tt PSRCHIVE} distribution (\cite{vdo12}).

Full details of the data acquistion, reduction, analysis and
interpretation will be shortly be presented elsewhere (Stairs et al.,
in prep.).  Here we present a subset of the results.

\begin{figure}[t]
\begin{center}
 \includegraphics[width=4in]{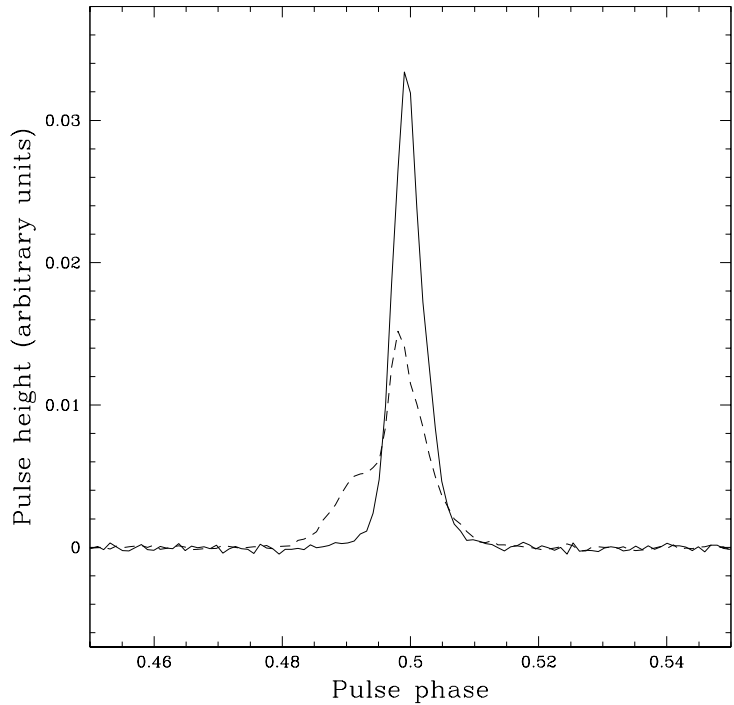} 
 \caption{Cumulative narrow (solid) and wide (dashed) profiles for
   data taken with Parkes on MJD 52466.}
   \label{profile}
\end{center}
\end{figure}

 Mode-changing is seen in many of the long observations, while other
days show only narrow or only wide profiles.  Fig.\,\ref{states}
summarizes the states seen on each day.  It is clear that the
transition rate of the mode changes varies as well as the average profile
shape.

Fig.\,\ref{profile} show the profiles observed with Parkes on MJD
52466.  These were obtained by summing, respectively, all the 10-s
sub-integrations labeled ``narrow'' or ``wide'' on that day.  The flux
density is uncalibrated.  The cumulative profile for each Parkes epoch
is well-described as a linear combination of these two extreme
profiles; the GBT data are similarly well-described using one day's
extrema.  This agreement with two extreme profiles argues strongly
against the precession model, in which one would expect smooth
changes.

The mode-changing transition rate has a relationship to the $\dot P$
cycle which may indicate that this quantity forms the second governing
variable in the chaos model; see Stairs et al. (in prep.) for a
complete discussion.  We advocate that other pulsars with known $\dot
P$ quasi-periodicities be carefully examined for evidence of
variable rates of mode-changing.

\section*{Acknowledgements}

IHS is supported by an NSERC Discovery Grant and by the Canadian
Institute for Advanced Research.  JvL received funding for this
research from the Netherlands Organisation for Scientific Research
(NWO) under project "CleanMachine" (614.001.301).    The Green Bank
Observatory is a facility of the National Science Foundation operated
under cooperative agreement by Associated Universities, Inc.   The
Parkes radio telescope is part of the Australia Telescope National
Facility which is funded by the Australian Government for operation as
a National Facility managed by CSIRO.  
We thank Ryan Hyslop, Jennifer Riley, Raymond Lum and Cindy Tam for
their work on earlier versions of the profile analysis.


\end{document}